\documentclass[journal]{IEEEtran}
\usepackage{amssymb}
\usepackage{amsmath}
\usepackage{graphicx}
\usepackage{cite}
\usepackage{citesort}

\usepackage{color}
\usepackage{multicol}
\usepackage{amsfonts}
\usepackage{bbm}
\usepackage{subfigure}
\usepackage{graphicx,epstopdf}
\usepackage{epsfig}	
\usepackage{cite,graphicx,amsmath,amssymb}
\usepackage{comment}
\usepackage{amsmath}
\usepackage{lipsum}
\usepackage{stfloats}
\usepackage{algorithm}
\usepackage{algorithmic}
\usepackage{bm}
\newlength{\halfpagewidth}
\divide\halfpagewidth by 2

\newtheorem{theorem}{\textbf{Theorem}}

\newtheorem{lemma}{\textbf{Lemma}}

\newtheorem{corollary}{\textbf{Corollary}}

\newtheorem{proof}{\textbf{Proof}}

\newtheorem{definition}{\textbf{Definition}}

\newtheorem{proposition}{\textit{Proposition}}

\makeatletter
\def\ScaleIfNeeded{%
\ifdim\Gin@nat@width>\linewidth \linewidth \else \Gin@nat@width
\fi } \makeatother

\begin{document}
%

\title{{A New Look at Physical Layer Security, Caching, and Wireless  Energy Harvesting for Heterogeneous Ultra-dense Networks}}
\author{Lifeng Wang,~\IEEEmembership{Member,~IEEE,}  Kai-Kit Wong,~\IEEEmembership{Fellow,~IEEE},  Shi Jin,~\IEEEmembership{Member,~IEEE,}\\
 Gan Zheng,~\IEEEmembership{Senior Member,~IEEE,}  and Robert W. Heath, Jr.,~\IEEEmembership{Fellow,~IEEE}
\thanks{L. Wang, and K.-K. Wong are with the Department of Electronic and Electrical Engineering, University College London, WC1E 7JE, London, UK (E-mail: $\rm\{lifeng.wang, kai$-$\rm kit.wong\}@ucl.ac.uk$).}
\thanks{S. Jin is with National Mobile Communications Research Laboratory, Southeast University, Nanjing 210096, China (Email: $\rm{jinshi}@seu.edu.cn$).}
\thanks{G. Zheng is with the Wolfson School of Mechanical, Electrical and Manufacturing Engineering, Loughborough University, Leicestershire, LE11 3TU, UK (Email: $\rm{g.zheng}@lboro.ac.uk$).}
\thanks{Robert W. Heath, Jr. is with the Department of Electrical and Computer Engineering, The University of Texas at Austin, Texas, USA (E-mail: $\rm{rheath}@utexas.edu$).}

}

\maketitle

\begin{abstract}

Heterogeneous ultra-dense networks enable ultra-high data rates and ultra-low latency through the use of  dense sub-6 GHz and millimeter wave (mmWave) small cells with different antenna configurations.  Existing work has widely studied spectral and energy efficiency in such networks and shown that high spectral and energy efficiency can be achieved. This article investigates the  benefits of heterogeneous ultra-dense network architecture from the perspectives of three promising technologies, i.e., physical layer security, caching, and wireless energy harvesting, and provides enthusiastic outlook towards application of these technologies in heterogeneous ultra-dense networks. Based on the rationale of each technology,  opportunities and challenges are  identified to advance the research in this emerging network.

\end{abstract}

\section{Introduction}
Future wireless networks such as 5G are designed to support increasing mobile traffic while reducing energy consumption. To meet these targets, new radio-access technologies such as  massive multiple-input multiple-output (MIMO) and millimeter wave (mmWave) will be key components of future wireless access~\cite{F_Boccardi_5G}. Meanwhile, the new use cases such as drone delivery, autonomous driving, and smart homes require ultra-high reliability and ultra-low latency, and {such requirements  entail deployment of ultra-dense low-power small cells, since user equipment (UEs) need to be much closer to the network.}  Moreover,  {since blockage can severely deteriorate mmWave transmissions, mmWave small cells need to be densely deployed in an attempt
to provide seemless coverage~\cite{H_zhang_2017}.}  Therefore, {next-generation wireless networks will accommodate network densification and a variety of radio-access technologies with different antenna configurations in sub-6 GHz and mmWave bands~\cite{dantong-survey}.}

Heterogeneous ultra-dense network  consisting of dense sub-6 GHz and mmWave small cells is  a key driver to fulfil critical requirements in terms of reliability and latency, and support enormous connectivity triggered by massive Internet-of-things (IoT) and cellular
vehicle-to-everything (C-V2X),  which is also spotlighted by the industry~\cite{Nokia_2016}.
{Existing contributions have widely investigated the spectral and energy efficiency of heterogeneous ultra-dense networks~\cite{Jeffrey_5G,M_Kamel_2017}. This article aims to provide a comprehensive overview of physical layer security, caching, and wireless energy harvesting in such networks, and deliver insightful guidelines for creating new efficient solutions for security, content delivery and energy.}  Fig. 1 illustrates a heterogeneous ultra-dense network architecture that has the capabilities of these new technologies, where physical layer security can overcome malicious eavesdropping, caching can offload the core network traffic, and wireless energy harvesting can prolong the lifetime of UE's battery.
\begin{figure}[t!]
\centering
\includegraphics[width=3.8 in,height=2.8 in]{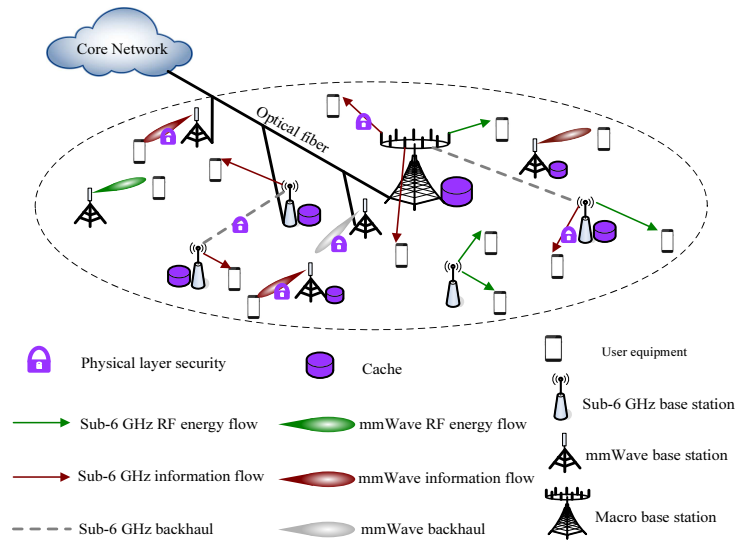}
\caption{Heterogeneous ultra-dense network that have the capabilities of physical layer security, caching, and wireless energy harvesting. In such networks, wired and wireless backhaul coexist.}
\label{WPT_UE}
\end{figure}

{{The remainder of this article is organized as follows. Section II identifies the key features of heterogeneous ultra-dense networks. Sections III, IV and IV illustrate physical layer security, caching and wireless energy harvesting in such networks, respectively, and provide new opportunities and challenges. Finally, conclusions are drawn in Section VI.}}

\section{Key Features in Heterogeneous Ultra-dense Networks}
In traditional heterogeneous networks (HetNets), the main difference between tiers is the level of transmit power.  The configuration of BSs in heterogeneous ultra-dense networks will be much diverse, with different tiers providing different levels of transmit power, antenna array gain, bandwidth and backhaul cost. The key features of future heterogeneous ultra-dense networks can be identified as follows:
\begin{itemize}
  \item \textbf{Channel Model:} Sub-6 GHz and mmWave links will coexist in heterogeneous ultra-dense networks. Compared to the high diffraction and penetration characteristics of sub-6 GHz links, the detrimental effect of blockage on mmWave links is significant, and mmWave links undergo more dramatic swings between line-of-sight (LoS) and non-line-of-sight (NLoS) due to high level of shadowing. { The coherence time in mmWave frequencies is about an order of magnitude shorter than that in sub-6 GHz frequencies  since the Doppler shift linearly scales with frequency.} Moreover, unlike rich scattering sub-6 GHz scenarios, there only exists limited spatial selectivity in mmWave environment, which makes traditional small-scale fading models invalid.

 \item \textbf{Frequency Band:} {In heterogeneous ultra-dense networks, multiple frequency bands will be leveraged, and UEs  may  experience different pathloss conditions,  since mmWave channels are more sensitive to blockages.}

 \item \textbf{BS Densification:} A very large number of small cells will be deployed in heterogeneous ultra-dense networks, to support huge amount of connectivity. In such networks, communication distances become much shorter than ever before, which reduces
     pathloss and thus UEs' transmit power during information transmission. Meanwhile, there are some unique features in such networks. For example, densifying the sub-6 GHz tier (usually interference-limited) leads to constant positive  downlink signal-to-interference-plus-noise ratio (SINR)~\footnote{The near-field pathloss exponent is larger than 2.} when all BSs with the same transmit power are active and none of interference mitigation approaches are utilized~\cite{Jeffrey_5G}. However, {densifying  the mmWave tier can significantly enhance the SINR, thanks to directional pencil-beams and its sensitivity of blockage~\cite{Tianyang_arxiv2014}.}

  \item \textbf{Beamforming/Precoding:} Each BS will be equipped with dozens or hundreds of antennas in an attempt to achieve large array gain or multiplexing gain. More importantly, mmWave BSs can pack more antennas than sub-6 GHz BSs because of shorter mmWave wavelengths. Due to hardware constraints, mmWave  beamforming/precoding may be hybrid analog/digitial, in contrast to the digital sub-6 GHz beamforming/precoding designs. Hence the array gains or multiplexing gains achieved by sub-6 GHz and mmWave BSs may be distinct.
  \item \textbf{Self Backhaul:}  It is challenging to let each backhaul link be fiber-optic in such networks, and a large number of small cells   are anticipated to access to the core networks via wireless backhaul. By leveraging massive MIMO and mmWave,  high-speed wireless backhaul links can be realized. In addition, multi-hop wireless self-backhauling is also required to enable flexible extension of the coverage range, which has been established in technical specification of 5G system by 3GPP~\cite{Caching_3GPP_2017}.
\end{itemize}

\section{Physical Layer Security-enabled Heterogeneous Ultra-dense Networks}
{Traditional cryptographic techniques may result in high-latency and communication failure in fast changing networks, due to higher-layer key distribution and management. Physical layer security is a low-complexity approach to keep messages confidential in the presence of malicious eavesdroppers~\cite{Zhu_Kit_Robert_2017}.} The principle of physical layer security is to exploit the randomness of wireless channels to achieve secrecy on physical layer. {In heterogeneous ultra-dense networks,  wireless channels are more diverse and random than ever before,  as indicated in Section II. Therefore, the application of physical layer security in such networks is promising.}  We will explain how heterogeneous ultra-dense networks can be powerfully combat malicious eavesdropping with the assistance of  BS densificaition and large antenna arrays.

{\textbf{BS Densificaition:}} One of the main benefits in heterogeneous ultra-dense networks is that by deploying a very large number of BSs, significant BS densification gains can be achieved to boost the spectral efficiency~\cite{Jeffrey_5G}, due to lower pathloss. In particular, only the active UEs can reap such benefits, and eavesdroppers that intend to overhear certain UEs' transmissions are not benefited but receive more inter-cell interference from interfering BSs. This can be understood by the fact that in the practical cellular networks, UEs are connected to the BSs based on specific user association rule such as maximum receive signal strength (max-RSS) and make handovers between adjacent BSs~\cite{dantong-survey}, however, eavesdroppers have no choice but to undergo arbitrarily varying channel conditions when intercepting a specific UE. The densities of BSs and active UEs determine the amount of BS densification gains and inter-cell interference. {The density of eavesdroppers determines the level of eavesdropping ability in the networks, i.e., there may exist stronger eavesdropping channels when more eavesdroppers are in  certain UEs' proximity.}

{\textbf{Large Antenna Arrays}:} Since eavesdroppers cannot obtain any transmit antenna array gains from multi-antenna BSs,  BSs equipped with large antenna arrays can provide legitimate UEs with large array gains and thus significantly enhance secrecy performance. The use of large antenna arrays also allows BSs or UEs to further cut transmit power, which in return reduces the received signal power of eavesdroppers. In heterogeneous ultra-dense networks, sub-6 GHz BSs and mmWave BSs may provide different levels of array gains based on different beamforming/precoding designs. Because BS can fit more mmWave antennas than sub-6 GHz antennas for a fixed antenna aperture, the mmWave beams may be much narrower than the sub-6 GHz counterpart, which degrades mmWave eavesdropping channels.

 {\textbf{Interference:}} In heterogeneous ultra-dense networks, enormous connectivity results in severe interference. For UEs, the obtained BS densification gains and array gains can well combat the interference. However, eavesdroppers can only alleviate the harm of interference by increasing their receive antenna array gains or colluding with each other. As mentioned before, an interesting feature of heterogeneous ultra-dense networks is that when densifying the sub-6 GHz tier in which all the BSs are assumed to use the same transmit power and near-field pathloss is larger than 2, the SINRs of UEs nearly remain unchanged\footnote{In reality, small cells may be more lightly loaded, and densifying the sub-6 GHz tier will improve the SINRs of UEs~\cite{Jeffrey_5G,Lifeng_JSAC_2017}.} but the SINRs of eavesdroppers will significantly decrease due to very strong interference. In this case, the SINRs of eavesdroppers tend to be zero as the density of sub-6 GHz BSs goes to infinity, and the SINRs of UEs can still be improved by obtaining more array gains from their associated BSs, which means that perfect secrecy is very likely to be achieved. Unlike sub-6 GHz tier, densifying the mmWave tier improves the SINRs of UEs,  i.e., the inter-cell interference power grows at a lower speed compared to the UEs' signal power, due to the fact that mmWave signals are transmitted via narrow beams and are susceptible to blockage. Densifying the mmWave tier will inflict more interference on eavesdroppers, which degrades the eavesdropping channels. Fig. 2 shows the effect of different densities of nodes  on the average transmission rate and average secrecy rate in a heterogeneous ultra-dense network where UEs can be connected to sub-6 GHz BSs or mmWave BSs under max-RSS based user association. It is seen that the average secrecy rate significantly increases with the density of mmWave BSs, and the average secrecy rate converges to the average transmission rate when the mmWave tier becomes ultra-dense, which indicates that little information will be overheard by eavesdroppers. The reason is that mmWave interference power can still be large when the mmWave tier is extremely dense, which deteriorates the eavesdropping channels. Moreover, the average transmission rate and average secrecy rate slowly increase with density of mmWave BSs after a critical point, due to the fact that ultra-dense mmWave tier can still be interference-limited~\cite{Tianyang_arxiv2014}. In addition, Fig. 2 shows that when the sub-6 tier is more dense, the average secrecy rate decreases because UEs will be more likely to select sub-6 GHz links with stronger signal strength and smaller bandwidth.
\begin{figure}[t!]
\centering
\includegraphics[width=3.5 in,height=2.8 in]{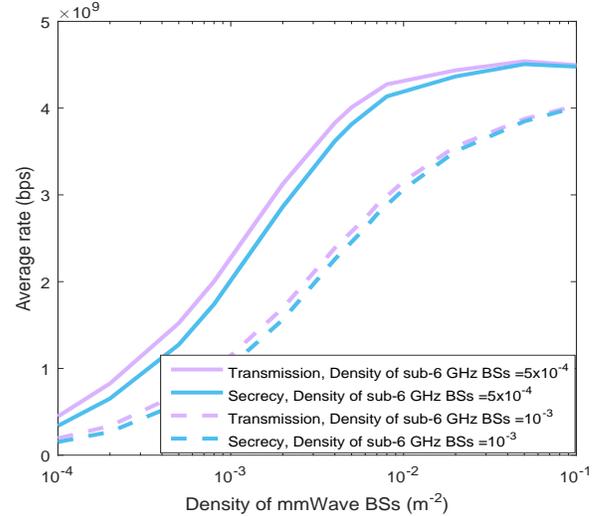}
\caption{Average secrecy rate under max-RSS based user association: Each sub-6 GHz BS has 16 antennas, and the main lobe gain, side lobe gain, and half power beamwidth of the mmWave BS antenna are $15~\mathrm{dB}$, $-15~\mathrm{dB}$, and $4.5^{o}$, respectively. The  sub-6 GHz carrier frequency is $1$ GHz and its pathloss exponent is $3.2$. The mmWave carrier frequency is 28 GHz and its LoS pathloss exponent is 2. The sub-6 GHz and mmWave bandwidths are 10 MHz and 1000 MHz, respectively. The radius of mmWave LoS region is 50 m. Each UE or eavesdropper has one single omnidirectional antenna. The sub-6 GHz BSs and mmWave BSs are assumed to use  same transmit power set as 30 dBm. Non-colluding eavesdropping scenario is considered and the density of eavesdroppers is 10$^{-3}$ m$^{-2}$.}
\label{cooperatove_caching}
\end{figure}

The aforementioned benefits have shown that heterogeneous ultra-dense networks with physical layer security can well preserve security and privacy of connectivity services. Moreover, advanced interference mitigation designs in heterogeneous ultra-dense networks may still need to be developed for security enhancement, particularly in sub-6 GHz tier which is usually interference-limited. It should be noted that whether the mmWave tier is noise-limited or interference-limited depends on the specific mmWave carrier frequency, mmWave beam pattern, blockage probability of the setting, and density of mmWave BSs. Thus, interference mitigation in mmWave tier may still be required when it is interference-limited~\cite{Tianyang_arxiv2014}. In addition, to create more flexible and scalable physical layer security designs in heterogeneous ultra-dense networks, researchers are encouraged to study new risks and security threats involving new applications and use cases such as critical infrastructure and industry processes.

\section{Cache-enabled Heterogeneous Ultra-dense Networks}
Video-based services and personal data storage applications have been instrumental for massive mobile traffic growth, and efficient content delivery is required in  future wireless networks. Recently, 3GPP has released the technical specification for services requirements in 5G system, in which 5G shall support content caching applications and operators are required to place the content caches close to UEs~\cite{Caching_3GPP_2017}. We will show that the content services will be delivered with very high reliability and very low latency in cache-enabled heterogeneous ultra-dense networks, where a very large number of BSs with caches have the ability of storing contents requested by UEs.

\begin{figure}[t!]
\centering
\includegraphics[width=3.5 in,height=2.8 in]{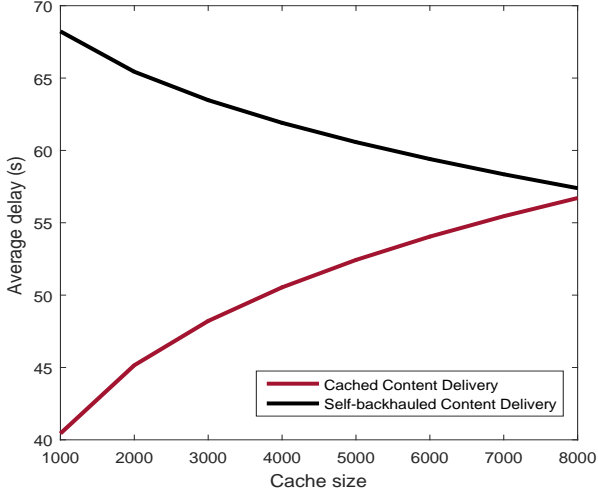}
\caption{Average delay for delivering a file in a two-tier self-backhauled heterogeneous ultra-dense network where each sub-6 GHz single-antenna  small BS (SBS) with finite cache size stores the most popular contents, and each sub-6 GHz massive MIMO aided macro BS (MBS) equipped with 128 antennas accesses to the core networks via optical fiber and delivers the non-cached contents to the SBSs by using zero-forcing transmission with equal power allocation. Max-RSS based user association is employed. The content library contains $10^5$ files and each file has 1 Gbit. The sub-6 GHz carrier frequency is 3.5 GHz, and bandwidths are orthogonally allocated to the cached and self-backhauled content delivery, which are 45 MHz and 55 MHz, respectively. The pathloss exponents of access link and backhaul link  are 3.0 and 2.6, respectively. The densities of SBSs and MBSs are $3\times 10^{-4}$ and $10^{-5}$, respectively. The typical macrocell load and small cell load are 10 and 5, respectively, and the minimum distance between SBS and MBS is assumed to be 5 m. The content popularity follows the Zipf distribution with the popularity skewness  0.7.}
\label{caching_latency}
\end{figure}

Cache-enabled BSs can deliver the cached contents to UEs at a short distance in heterogeneous ultra-dense networks, which significantly reduces the latency. The connectivity in such networks is highly reliable at a high-speed data rate, owing to the BS densification gains,  array gains, and large mmWave bandwidths.  Therefore, in heterogeneous ultra-dense networks, whether UEs' requested contents are cached by the serving  BSs or not dominates the performance of the network. {As indicated in Fig. \ref{caching_latency},  the average delay for cached content delivery is lower than self-backhauled content delivery. The average delay for cached content delivery increases with increasing the cache size.  In contrast, the average delay for self-backhauled content delivery decreases with increasing the cache size. The reason is that larger cache size results in higher hit probability, and more SBSs can provide cached content delivery, which results in more inter-SBS interference over the  frequency band allocated to the cached content delivery, and less inter-SBS interference over the frequency band allocated to the self-backhauled content delivery.}

Hit probability is an important metric to characterize the probability that a requested content file is stored by an arbitrary BS. A lower hit probability means that more content services are provided via wired/wireless backhaul. Different content placement strategies result in different hit probabilities, since the content placement mechanism  determines whether a content should be cached by a BS or not. Content placement mechanism is mainly designed based on content popularity, and caching the most popular contents (MPC) at each BS is an intuitive and  low-complexity approach, which achieves the highest hit probability for fixed cache size.  However, MPC caching may not be optimal in dense HetNets, and existing contributions have attempted to propose some optimal content placement solutions~\cite{ZCH-COO}, to maximize the probability that the requested content files are not only cached by the serving BSs but also successfully delivered to the associated UEs. In light of the aforementioned heterogeneous ultra-dense system features, caching design in such networks may need to take into account the
 following important aspects:
\begin{itemize}
 \begin{figure}[t!]
\centering
\includegraphics[width=3.5 in,height=2.8 in]{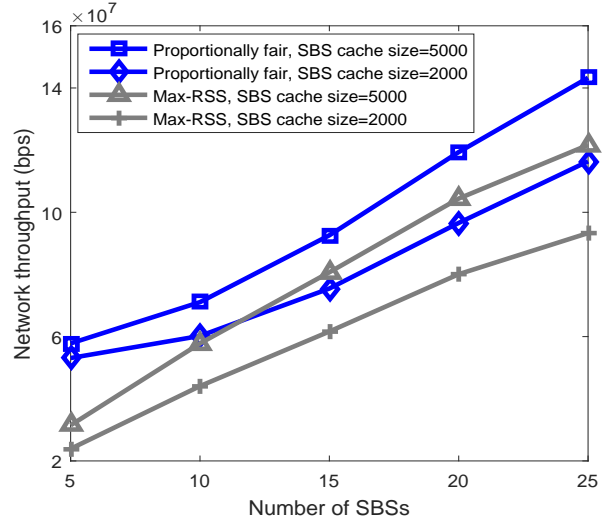}
\caption{Effect of different user association algorithms on network throughput: Consider  proportionally fair based~\cite{QY12} and  max-RSS based user association schemes in a sub-6 GHz HetNet consisting of one macro BS (MBS) and multiple small BSs (SBSs). SBSs are randomly located in this macrocell geographical area, and the macrocell radius is 300 m. The sub-6 GHz bandwidth is 10 MHz, and the number of UEs is 40. All the SBSs use same transmit power and the transmit power of MBS and SBS are 40 dBm and 30 dBm, respectively. The content popularity follows the Zipf distribution with the popularity skewness  0.8. Each content has unit size, the content library size is 10$^{5}$, and the MBS's cache size is 7000. The pathloss  between MBS and UE is $128.1 + 37.6{\log _{10}}d\left( {\mathrm{km}} \right)$, and between SBS and UE is $140.7 + 36.7{\log _{10}}d\left( {\mathrm{km}} \right)$, respectively.}
\label{caching_UA}
\end{figure}
  \item \textbf{User Association:} In contrast to the conventional user association schemes that mainly aim to improve spectral and energy efficiency, UEs may be given priority to be associated with the nearby BSs that have cached the requested contents, to reduce backhaul cost and latency. The reason is that under the conventional user association schemes such as max-RSS,  UEs may be connected to the BSs without caching the requested contents, which will result in more backhaul and inevitably increase the latency and the backhaul cost. Therefore, content-aware user association needs to be developed. In addition, the throughput and energy efficiency differences  between sub-6 GHz link and mmWave link will play a key role in determining which type of link to deliver the same contents, since spectral and energy efficiency are two of the key performance indicators in 5G. Fig. 4 gives an example to illustrate the effect of user association on the network throughput under MPC caching. It is observed that different user association schemes achieve different levels of network throughput, and deploying more SBSs can improve the network throughput, due to BS densification gains. Expanding the cache size improves network throughput since the contents requested by UEs are more likely cached at nearby BSs, which means that the core network traffic can be greatly offloaded. The proportionally fair based user association outperforms  max-RSS, which can be explained by the fact that in conventional  max-RSS based user association scheme, UEs only select the strongest channels and the effect of cache hit is ignored.

  \item \textbf{Cache Size:} BSs in different tiers may have different cache sizes, and such heterogeneity also has a big impact on efficient content delivery. In cache-enabled heterogeneous ultra-dense networks, the appropriate cache size of a tier heavily depends on the density of BSs, latency and backhaul cost, i.e., tiers with low latency and low backhaul cost may only need to fit small caches, since the uncached  contents can be obtained via wired/wireless backhaul, however, tiers with high latency or high backhaul cost shall have more BSs with large caches.

  \item \textbf{Dual Connectivity:}  UEs may have the ability of dual radio capabilities to be connected to multiple BSs in different types of tiers at the same time, e.g., in 5G systems, a UE can operate at sub-6 GHz band and mmWave band simultaneously~\cite{Caching_3GPP_2017}. In such cases, content placement optimization designs may be more complicated, since the associated BSs can cooperatively serve a UE. As shown in Fig. 5, there are two types of cooperative caching, i.e., \textbf{Type I:} the associated BSs transmit the same content file to a UE, to achieve transmit diversity; and \textbf{Type II:} each associated BS caches partition of a content file and UE receives all the partitions from its associated BSs, to achieve content diversity. It should be noted that for \textbf{Type II} cooperative caching, the partition size of a content may be different between the associated sub-6 GHz BS and mmWave BS, since mmWave BS with gigahertz bandwidth supports much higher data rate.
      \begin{figure}[t!]
\centering
\includegraphics[width=3.5 in,height=2.8 in]{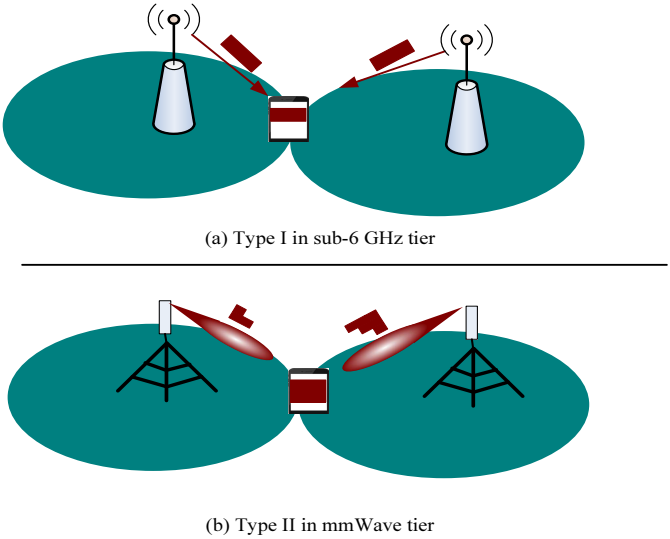}
\caption{Cooperative caching strategies (Note that Type I and Type II may coexist in the same tier).}
\label{cooperatove_caching}
\end{figure}
   \item \textbf{Integrated Access and Backhaul:}  Content placement has a significant effect on the backhaul load, due to the fact that uncached contents have to be obtained via backhaul. The self-backhaul feature of heterogeneous ultra-dense networks creates a need for integrating access and backhaul links to improve the efficiency of content delivery. Therefore, load balancing problem in cache-enabled heterogeneous ultra-dense networks with integrated access and backhaul links needs to be addressed. Since the access and backhaul links may be operated on the same or different frequencies, spectrum allocation solutions for integrated access and backhaul links are also required to improve spectral efficiency in content delivery.

\end{itemize}

Cache-aware resource allocation optimization methods can  be developed to enhance the efficiency of content delivery. However, it is challenging to provide the holistic design in cache-enabled heterogeneous ultra-dense networks with very large numbers of BSs in practice, because it is hard to obtain all the BSs' channel station information (CSI). The application of mean-field theory could be a tractable approach, which allows BSs to make their own decision based on local CSI while abstracting other BSs' strategies using a mean-field~\cite{Samarakoon_2016}. To date, there are few research results available yet for presenting resource allocation designs in cache-enabled heterogeneous ultra-dense networks, and researchers are encouraged to pay attention to this area.

Security and privacy are of great importance in cache-enabled heterogeneous ultra-dense networks, since the content delivery needs to be protected from being intercepted or attacked by illegitimate UEs. Meanwhile, the growing demand for content services such as video puts pressure on network management when implementing traditional content encryption schemes, and the multi-hop characteristic of backhaul links in heterogeneous ultra-dense networks gives rise to more complexity in higher-layer key distribution and management. More importantly, the public cares deeply about their privacy such as locations and browsing history when requesting content services. Hence new security designs need to be developed for facilitating network management under various security concerns. As suggested in Section III, physical layer security may be a cost-effective security solution for dealing with security issues in cache-enabled heterogeneous ultra-dense networks, particularly securing access and backhaul links. Currently, the research on securing cache-enabled heterogeneous ultra-dense networks is in its infancy, and more research efforts need to be made.

\section{Wireless Energy Harvesting-enabled Heterogeneous Ultra-dense Networks}
Wireless energy harvesting is a much more controllable approach to prolong the lifetime of UEs or devices in IoT, compared to the traditional renewable energy harvesting that heavily depends on the conditions of the environments. The rationale behind it is that {radio-frequency (RF)} energy can be harvested via microwave radiation  in sub-6 GHz or mmWave frequency. RF-to-DC conversion  and pathloss are two key factors that influence the efficiency of wireless energy harvesting. {Sophisticated rectifier circuit hardware needs to be designed to achieve high RF-to-DC conversion efficiency such that more RF energy can be harvested  during RF-to-DC conversion.} Pathloss  imposes a limit on the received signal power, and thus the distance between UEs and BSs (i.e., RF energy sources) cannot be too long since the amount of received RF energy is required to be large enough to activate the harvesting circuit. We believe that heterogeneous ultra-dense networks provide a wealth of opportunities for wireless energy harvesting based on the following advantages:
\begin{itemize}
  \item \textbf{Large Antenna Arrays:} The use of large antenna array forms very sharp signal beams, to achieve significant array gains and redeem pathloss. The obtained array gains from sub-6 GHz BSs and mmWave BSs may be different, and mmWave BSs may provide very large array gains given the fact that very shorter mmWave wavelengths enable BSs to fit a very large number of mmWave antennas.
  \item \textbf{Network Densification:} Compared to the today's networks, UEs will experience less pathloss in heterogeneous ultra-dense networks, which is of great importance for wireless energy harvesting. Moreover, interference power from enormous access and self-backhaul links could also be harvested by UEs.

  \item \textbf{Dual Connectivity: } Since a UE can be associated with multiple BSs (multiple RF energy sources) at the same time, it can be cooperatively powered by its serving BSs, to obtain more array gains and BS densification gains. Therefore, dual connectivity can be an appealing approach to quickly recharge high-power UEs via BS cooperation. {It should be noted that dual connectivity for wireless power transfer will inevitably result in more energy loss since different serving BSs undergo different pathloss conditions and far-away serving BSs have higher pathloss. Therefore, power cost issue needs to be addressed in realistic networks.}
\end{itemize}

There are two types of RF energy sources in wireless energy harvesting-enabled heterogeneous ultra-dense networks, namely the associated BSs (as dedicated RF energy sources) that provide the directed power transfer, and the remaining BSs that act as ambient RF energy sources. Since the associated BSs  deliver much higher amount of RF energy to the UEs via directed beams than other BSs~\cite{Lifeng_JSAC_2017}, high-power UEs have to be powered by their associated BSs. Therefore, to maximize the directed transferred energy, high-power UEs are associated with BSs under max-RSS based user association. Considering the fact that enormous connectivity results in high interference power in heterogeneous ultra-dense networks,  a lower-power UE  or device in IoT can harvest such energy and thus it is unnecessary to be associated with a specific BS. In addition, a high-power UE with dual-mode may ask for directed power transfer from a sub-6 GHz BS or mmWave BS. Fig. \ref{WPT_UE} shows the effect of BS density on power transfer user association. It is observed that UEs are more likely to select mmWave links to deliver RF energy when mmWave tier is ultra-dense, and densifying the sub-6 GHz tier means that the probability for UEs being powered by sub-6 GHz BSs increases. In addition, adding more  antennas can improve sub-6 GHz power transfer association probability because of more array gains.
\begin{figure}[t!]
\centering
\includegraphics[width=3.5 in,height=2.8 in]{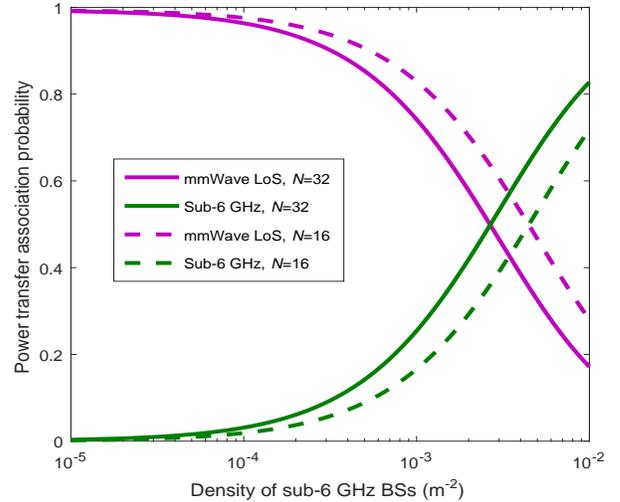}
\caption{Power transfer association probability under max-RSS based user association: The density of mmWave BSs is  $10^{-2}~\mathrm{m}^{-2}$. Each sub-6 GHz BS is equipped with $N$ antennas, and the main lobe gain, side lobe gain, and half power beamwidth of the mmWave BS antenna are $18~\mathrm{dB}$, $-2~\mathrm{dB}$, and $10^{o}$, respectively. The  sub-6 GHz carrier frequency is $1.5$ GHz and its pathloss exponent is $2.7$. The mmWave carrier frequency is 28 GHz and its LoS pathloss exponent is 2. The mmWave LoS probability function is $e^{-d/141.4}$ with the distance $d$. Each UE has one single sub-6 GHz antenna, and the main lobe gain, side lobe gain, and half power beamwidth of its mmWave antenna are $10~\mathrm{dB}$, $-10~\mathrm{dB}$, and $45^{o}$, respectively. The sub-6 GHz BSs and mmWave BSs are assumed to use  same transmit power.}
\label{WPT_UE}
\end{figure}

In wireless powered heterogeneous ultra-dense networks, UEs first harvest the RF energy from their associated BSs, and then utilize the harvested energy to transmit messages.  In such networks, user association may need to be designed by considering the below aspects:
\begin{itemize}
  \item \textbf{Coupled:}   In this case,  UEs are associated with the same BSs in both downlink and uplink phases, and it could be downlink-based or uplink-based. The downlink-based user association can maximize the harvested energy, and the uplink-based user association can strengthen the received information signal power for energy saving. Therefore, it is important to justify the coupled association policy based on downlink or uplink.
  \item \textbf{Decoupled:} In this case, UEs can be associated with different BSs in downlink and uplink phases. Although such user association mechanism can enable UEs to select the best links for energy harvesting and information transmissions, channel reciprocity in massive MIMO or mmWave links will be lost. Therefore, link budget issues need to be carefully addressed.
\end{itemize}
In addition, the ratio of BS density to active UE density is one of the key system design parameters, which determines the amount of obtained BS densification gains in order to alleviate the harm of uplink interference, since dense active UEs can still result in high uplink interference in wireless powered dense networks. Moreover, to avoid using complicated and time-consuming interference mitigation schemes, one appealing alternative could be that UEs  harvest RF energy from sub-6 GHz BSs or mmWave BSs, but only choose mmWave links to deliver information messages, considering the fact that the uplink of mmWave tier tends to be noise-limited when active UEs are not extremely dense~\cite{Singh_2015}.

Downlink simultaneous wireless information and power transfer (SWIPT) also needs to be investigated in heterogeneous ultra-dense networks. There are two commonly-considered SWIPT protocols, namely power splitting and time switching~\cite{Rui_Zhang_2013}, and it would be interesting to study which SWIPT protocol is more suitable in such networks. Moreover, rectifier circuit at different frequencies may achieve different levels of  RF-to-DC conversion efficiency, and it is essential to select appropriate rectifier circuit in multi-band heterogeneous ultra-dense networks.

\section{Conclusions}
Heterogeneous ultra-dense networks fuse various technologies including ultra-dense small cells, massive MIMO and mmWave. The unique features of such network architecture provide physical layer security, caching, and wireless energy harvesting with new opportunities.  We have illustrated the benefits of using physical layer security, caching, and wireless energy harvesting in heterogeneous ultra-dense networks, and identified technical challenges, respectively. Since security, content services, and energy are of paramount importance (e.g., 5G service requirements~\cite{Caching_3GPP_2017}), the new solutions introduced by this article can help engineers  form the basis of efficient future networks.

\bibliographystyle{IEEEtran}

\begin{IEEEbiography}
{Lifeng Wang} [M] is the postdoctoral research fellow in the Department
of Electronic and Electrical Engineering,
University College London (UCL). He received the
Ph.D. degree in Electronic Engineering at
Queen Mary University of London  in  2015. His research interests include massive MIMO, millimeter wave, dense HetNets, edge caching,  physical-layer security, and wireless energy harvesting. He received the Exemplary Reviewer Certificate and the Exemplary Editor Certificate of the IEEE Communications Letters in 2013 and 2016, respectively.
\end{IEEEbiography}

\begin{IEEEbiography}
{Kai-Kit Wong} [F] is Professor of Wireless Communications at the Department of Electronic and Electrical Engineering, University College London, United Kingdom. He received the BEng, the MPhil, and the PhD degrees, all in Electrical and Electronic Engineering,
from the Hong Kong University of Science and Technology, Hong Kong, in 1996, 1998, and 2001, respectively. He is Fellow of IEEE and IET. He is Senior Editor of IEEE Communications Letters and also IEEE Wireless Communications Letters.
\end{IEEEbiography}

\begin{IEEEbiography}
{Shi Jin} [SM] is Professor at the faculty of the National Mobile Communications Research Laboratory, Southeast University. His research interests include 5G and beyond, random matrix theory, and information theory. He serves as an Associate Editor for the
IEEE Transactions on Wireless Communications. He has been awarded the 2011 IEEE Communications Society Stephen O. Rice Prize Paper Award in the field of communication theory, and the 2016 GLOBECOM Best Paper Award.
\end{IEEEbiography}

\newpage

\begin{IEEEbiography}
{Gan Zheng} [SM] is a Senior Lecturer in Wolfson School of Mechanical, Electrical and Manufacturing Engineering, Loughborough University, UK. He received the BEng and MEng from Tianjin University, China, and the PhD degree from The University of Hong Kong in 2008. His research interests include edge caching, full-duplex radio, wireless power transfer, and physical-layer security. He received the 2013 IEEE Signal Processing Letters Best Paper Award, and the 2015 GLOBECOM Best Paper Award.
\end{IEEEbiography}

\begin{IEEEbiography}
{Robert W. Heath Jr.} [F] is a Cullen Trust for Higher Education Endowed Professor in the Department of ECE at The University of Texas at Austin and Director of UT SAVES. He has received several awards including the 2017 Marconi Prize Paper Award and the 2017 EURASIP Technical Achievement Award. He is a licensed Amateur Radio Operator, a registered Professional Engineer in Texas, and a Private Pilot.
\end{IEEEbiography}

\end{document}